\documentclass[
 aip,
 amsmath,amssymb,
 floatfix,
 reprint,
]{revtex4-1}

\usepackage{graphicx}% Include figure files
\usepackage{dcolumn}% Align table columns on decimal point
\usepackage{bm}% bold math

\usepackage[utf8]{inputenc}
\usepackage[T1]{fontenc}
\usepackage{mathptmx}
\usepackage{etoolbox}
\usepackage[version=3]{mhchem} 
\usepackage{wrapfig}
\usepackage[export]{adjustbox}

\makeatletter
\def\@email#1#2{%
 \endgroup
 \patchcmd{\titleblock@produce}
  {\frontmatter@RRAPformat}
  {\frontmatter@RRAPformat{\produce@RRAP{*#1\href{mailto:#2}{#2}}}\frontmatter@RRAPformat}
  {}{}
}%
\makeatother
\begin{document}

\preprint{AIP/123-QED}

\title[Below the Surface: Unraveling the Intricacies of
the Nonlinear Optical Properties of Aluminum through Bound Electrons]{Below the Surface: Unraveling the Intricacies of the Nonlinear Optical Properties of Aluminum through Bound Electrons}
% Force line breaks with \\
\author{M. Scalora}
\affiliation{Aviation \& Missile Center, US Army CCDC, Redstone Arsenal, AL 35898-5000, USA}

\author{K. Hallman}
\affiliation{PeopleTec, Inc. 4901-I Corporate Dr., Huntsville, AL 35805, USA}

\author{S. Mukhopadhyay}
\affiliation{Department of Physics, Universitat Politècnica de Catalunya, Rambla Sant Nebridi 22, 08222 Terrassa (Barcelona), Spain}

\author{S. Pruett}
\affiliation{AFROTC Detachment 005, Auburn University, 243 Nichols Center, Auburn, AL 36849}

\author{D. Zappa}
\affiliation{Department of Information Engineering – University of Brescia, Via Branze 38, 25123 Brescia, Italy}

\author{E. Comini}
\affiliation{Department of Information Engineering – University of Brescia, Via Branze 38, 25123 Brescia, Italy}

\author{D. de Ceglia}
\affiliation{Department of Information Engineering – University of Brescia, Via Branze 38, 25123 Brescia, Italy}

\author{M. A. Vincenti}
\affiliation{Department of Information Engineering – University of Brescia, Via Branze 38, 25123 Brescia, Italy}

\author{N. Akozbek}
\affiliation{US Army Space \& Missile Defense Command, Tech Center, Redstone Arsenal, AL 35898 USA}

\author{J. Trull}
\affiliation{Department of Physics, Universitat Politècnica de Catalunya, Rambla Sant Nebridi 22, 08222 Terrassa (Barcelona), Spain}

\author{C. Cojocaru}
\altaffiliation{Corresponding author}
\affiliation{Department of Physics, Universitat Politècnica de Catalunya, Rambla Sant Nebridi 22, 08222 Terrassa (Barcelona), Spain}
\email{crina.maria.cojocaru@upc.edu}

\begin{abstract}
By uncovering novel aspects of second harmonic generation in aluminum we show that there are unusual and remarkable consequences of resonant absorption, namely an unexpectedly critical role that bound electrons play for light-matter interactions across the optical spectrum, suggesting that a different basic approach is required to fully explain the physics of surfaces. We tackle an issue that is never under consideration given the generic hostile conditions to the propagation of light under resonant absorption. Unlike most noble metals, aluminum displays Lorentz-like behavior and interband transitions centered near 810 nm, thus splitting the plasmonic range in an atypical manner and setting its linear and nonlinear optical properties apart. Studies of aluminum nanostructures having complex topologies abound, as do reported inconsistencies in the linear spectral response of surface plasmons and harmonic generation. Our experimental observations of second harmonic generation from aluminum nanolayers show that bound electrons are responsible for a unique signature neither predicted nor observed previously: a hole in the second harmonic spectrum. A hydrodynamic-Maxwell theory explains these findings exceptionally well and becomes the basis for renewed studies of surface physics. 

\end{abstract}




\maketitle

\section{Introduction}\label{sec1}

Plasmonics is concerned mostly with the interaction of light with free charges on conductive surfaces. Historically, noble metals like gold (Au) and silver (Ag) have been the preferred choices because of their relatively low losses in the visible and near-infrared (IR) ranges. The push toward the ultraviolet (UV) and beyond calls for additional studies of Au and Ag and alternative materials to determine their viability, a search that naturally highlights aluminum (Al) \cite{bib1}. Al is relatively inexpensive and abundant; it is stable with a robust spectral response in the UV range and is compatible with metal-oxide-semiconductor technology. While in some studies Al has been reported to outperform Ag in the visible range due to its seemingly superior surface and interface properties \cite{bib2,bib3}, most studies are either relegated to the UV range, or have been conducted at a single carrier wavelength\cite{bib4,bib5,bib6} with models based exclusively on free electron dynamics that lead to results inconsistent with expectations \cite{bib4}.

Unlike most noble metals, Al displays a broad and large absorption resonance near 810 nm, which is thought to relegate true plasmonic behavior to below 500 nm and above 1000 nm. As a result, there is a natural predisposition to avoid the resonance, to circumvent supposed detrimental effects of absorption, and to make approximations that leave that wavelength range unexplored and not well-understood. Here we set out to experimentally and theoretically study the nonlinear response of Al nanolayers of various thicknesses across the UV to the mid-IR, with the aid of a unique hydrodynamic-Maxwell model that accounts for linear and nonlinear material dispersions, as well as second and higher order surface and volume nonlinear sources. Our results suggest that bound charges play an unexpected, previously unknown yet outsized role in second harmonic generation (SHG), pointing to the fact that predictions based solely on the free electron model do not adequately describe light-matter interactions, and that the absorption resonance plays a pivotal role well beyond presupposed limits. Based on the results outlined below, we are led to the following conclusions: (1) Al does not behave like an ordinary free electron system, with repercussions to both linear and nonlinear light-matter interactions; (2) the influence of bound electrons should not be excluded or discounted even at wavelengths where pure plasmonic behavior is expected. The presence of bound electrons in the dynamics is necessary to conserve overall energy and momentum; (3) an open channel of energy flow in the bound electron dynamics can either act like a controlling catalyst or completely dominate the interaction, rendering the free-electron approach inadequate; (4) in a broad sense and beyond the specific material, there is a need to reevaluate the dynamical aspect of surface interactions beyond the simplest assumptions. 

\section{Outline of the theoretical model}\label{sec2}

\begin{figure*}[!ht] 
\centering
\includegraphics[width=.9\textwidth]{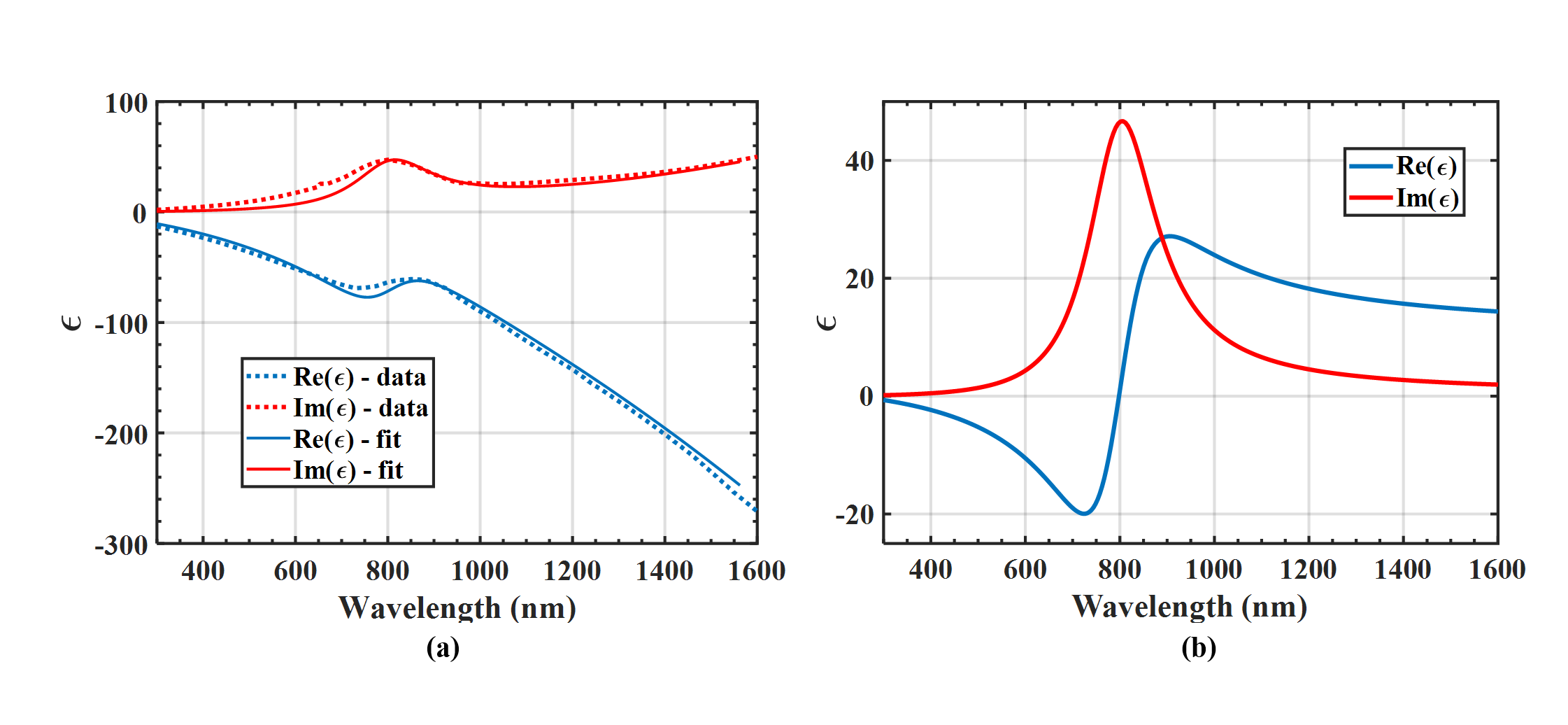}
\caption{(a) Total complex local dielectric response of aluminum, as reported in reference \cite{bib8} (dotted curves), and as fit by a combined Drude and Lorentz oscillator species (solid curves). (b) The isolated Lorentzian resonance peaked near 810 nm. Its influence may be seen to extend across the entire range shown.}\label{fig1}
\end{figure*}

The basic optical properties of Al and its band structure were first outlined in some detail in reference \cite{bib7}. More accurate, tabulated ellipsometric Al data has since become available \cite{bib8}. Most current studies of aluminum-based systems are done in the context of multi-dimensional nanostructured metasurfaces, for example nanopillars, apertures, holes, and other shapes, which are generally thought to enhance the plasmonic response, and thus overall performance \cite{bib9,bib10,bib11,bib12,bib13,bib14,bib15,bib16,bib17,bib18,bib19,bib20,bib21,bib22,bib23}. Typical examples of enhanced performance may relate to plasmon line shape narrowing \cite{bib21}, the enhancement of third harmonic generation (THG) from single hole apertures \cite{bib15}, and a doubly resonant nano-array to capture the resonant response of both pump and SH light \cite{bib19,bib20}. In all cases illumination is chosen to scrupulously avoid tuning the pump or any generated harmonic near the absorption resonance.  In Fig.\ref{fig1}(a) we plot the complex dielectric response of Al as reported in Ref. \cite{bib8}, together with the Drude-Lorentz functional fit that we use in our simulations. The dielectric function has the following scaled form: $\epsilon(\omega) = 1-\frac{\omega^2_{pf}}{\omega^2+i\gamma_f \omega}-\frac{\omega^2_{pb}}{\omega^2-\omega^2_{0b}+i\gamma_{0b} \omega}$. The scaled frequency $\omega = \frac{1}{\lambda}$ where ${\lambda}$ is expressed in microns; $\omega_{pf} = 10.5$ and $\omega_{pb} = 4$ are the respective plasma frequencies; $\gamma_f = 0.105$  and $\gamma_b = 0.315$  are the corresponding damping coefficients, and $\omega_{0b} = 1.24$ is the Lorentzian resonance frequency. The absorption resonance depicted in Fig.\ref{fig1}(b) seemingly splits the plasmonic range in two distinct regions. However, a comparison between Fig.\ref{fig1}(a) and \ref{fig1}(b) suggests that it is difficult to identify a region where the response is due solely to free electrons.

\begin{figure*}[!ht] 
\centering
\includegraphics[width=.9\textwidth]{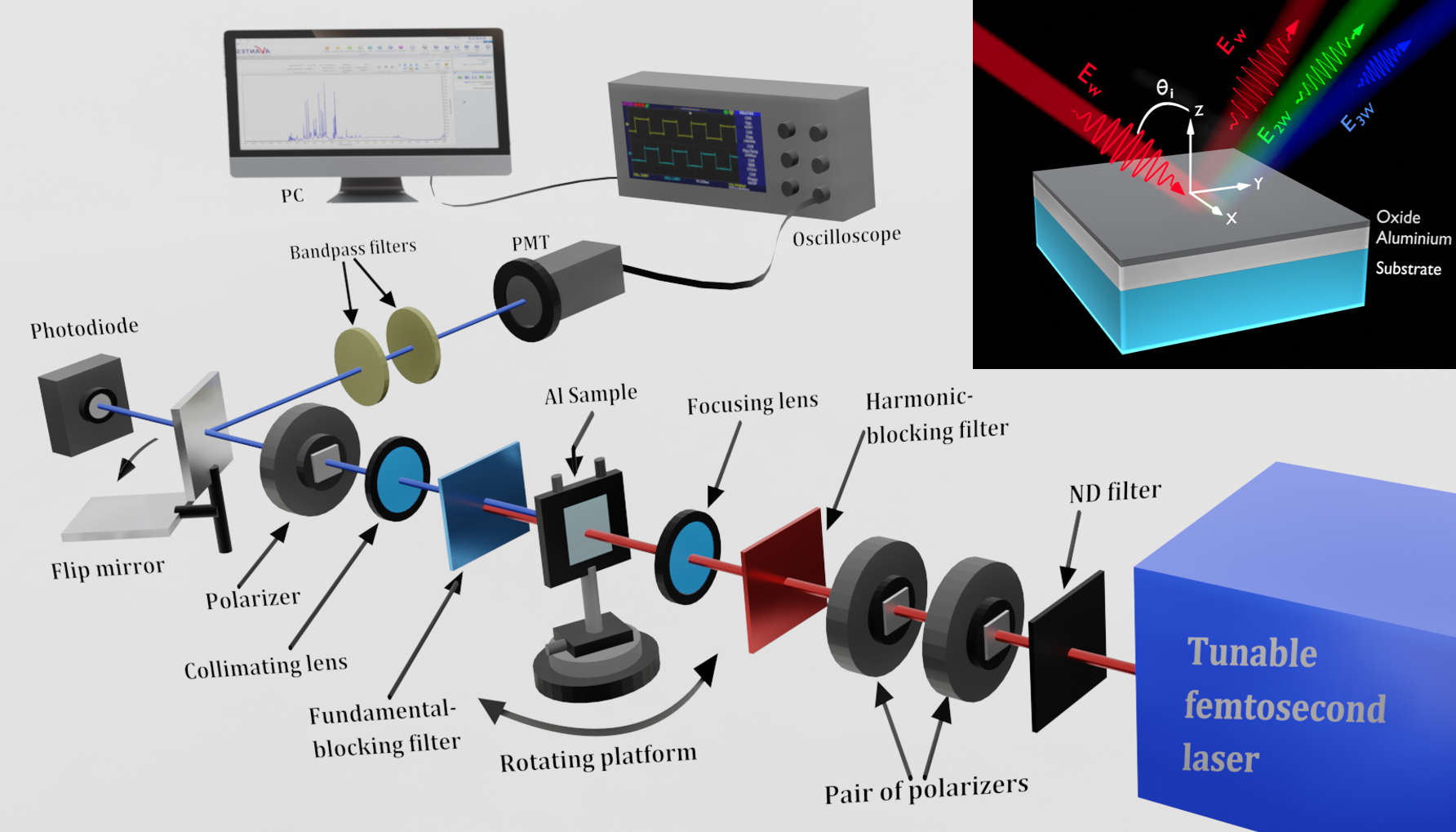}
\caption{Schematics of the setup and the aluminum layer deposited on a glass substrate (top right inset). Each component and its function are indicated in the figure.}\label{fig2}
\end{figure*}

In all referenced literature, the absorption resonance is simply noted but otherwise characterized as a mere limitation on the range of the plasmonic response, while the spectral regions above 1000 nm and below 500 nm are generally assumed to take on a Drude-like form of the type $\epsilon(\omega) = \epsilon_\infty - \frac{\omega^2_{pf}}{\omega^2+i\gamma_f \omega}$ , (e.g., reference \cite{bib13}), where $\epsilon_\infty$ is a constant that can either raise or lower the Drude curves, albeit without changing their curvatures. This apparent neglect may explain some of the discrepancies discussed in reference \cite{bib13}, which appear to indicate consistent redshifts between 50 nm and 100 nm in the theoretical predictions compared to experimental observations of the plasmonic resonance. The absorption resonance may in fact push its influence well into the visible and IR ranges. Put another way, the electrodynamics that takes place near the absorption resonance and its possible consequences to nonlinear interactions and plasmonic behavior are generally ignored. For example, the full Lorentzian function adds several basic ingredients that the introduction of a mere constant cannot: curvature and anomalous dispersion, and a response that simultaneously preserves linear and nonlinear dispersions and their amplitudes at the pump and its harmonic wavelengths. 

Based on the free electron model, there was expectation that at 810 nm Al should yield more SH light compared to Ag. However, experiments showed that the opposite occurs\cite{bib4}. Additional results also point to a large discrepancy in the magnitudes of the usual $a$, $b$, and $d$ coefficients used in the definition of the effective surface (longitudinal and transverse) and bulk
second order susceptibilities $\chi^{(2)}$, derived from the free electron model \cite{bib6}:

\begin{equation}
\begin{split}
     \chi^{(2)}_{\perp,surface} = a(\omega) \frac{\epsilon(\omega) - 1}{16 \pi m \omega^2}; \\
    \chi^{(2)}_{\parallel,surface} = b(\omega) \frac{\epsilon(\omega) - 1}{16 \pi m \omega^2}; \\
    \chi^{(2)}_{bulk} = d(\omega) \frac{\epsilon(\omega) - 1}{16 \pi m \omega^2}
\end{split}
\label{eqn:1}
\end{equation}

According to theory the magnitudes of $b$ and $d$ are expected to be of order unity. Instead, measurements fit the theory only if they are taken to be zero \cite{bib24}. The authors speculated that the effects of resonant absorption could be incorporated by merely adding the full linear dielectric function in the above definitions of $\chi^{(2)}$ \cite{bib4,bib6,bib24}. However, the authors acknowledged that these modifications to the theory were not only insufficient to explain the observed discrepancies, but left open the possibility that additional, unspecified physics might underlie these unexpected findings. Therefore, the notion that nonlinear second order contributions that might arise directly from bound electrons themselves was not entertained and continues to be missing in most theoretical treatments of centrosymmetric material nonlinearities. In contrast, we meticulously, consistently, and accurately model the material response of metals, semiconductors, and conductive oxides alike by making sure that linear and nonlinear dispersions are accounted for, with agreement between theoretical predictions and experimental results (see references \cite{bib24,bib25,bib26,bib27,bib28,bib29,bib30,bib31} and citations therein.)

A proper treatment of linear and nonlinear dispersions is therefore necessary if nonlinear refraction, absorption, frequency conversion, down-conversion and surface light-matter interactions are to be accurately predicted at the pump and its harmonic wavelengths. Noble metals, semiconductors and conductive oxides like silicon (Si), gallium arsenide (GaAs), and indium tin oxide (ITO) display absorption resonances in the UV range, where one usually finds some of the generated harmonic signals and where appropriate dispersion functions must be assigned. However, the most prominent feature in Al, the absorption resonance peaked near 810 nm, is inside the range where one generally tunes the pump field. It then stands to reason that both SHG and higher harmonics will be increasingly more sensitive to the nonlinear response of bound electrons as the pump is tuned across the resonance.

\section{Experimental set-up and sample description}\label{sec3}

Aluminum layers 20 nm and 40 nm thick were deposited by means of DC sputtering onto D263 Shott glass substrate (Fig.\ref{fig2} inset) (Ar plasma, DC power 50 W, pressure 4.4 Pa, deposition rate 0.22 nm/s). To cover the wavelength ranges required for this study we performed two sets of measurements using two different laser systems that yielded similar results in their range of overlap. The schematic setup is shown in Eq.\ref{eqn:2}. The first set of measurements was performed using a Ti: Sapphire laser tunable around 800 nm, emitting pulses of 150 fs at a 76 MHz repetition rate, with peak power densities (when the pump was focused on the sample) between 4 and 5 GW/cm\textsuperscript{2}. For the second set of measurements, we used an amplified Ti: Sapphire laser system pumping an optical parametric amplifier (OPA) that generates tunable 100 fs pulses at a 1 kHz repetition rate. We tuned the pump in the range of 600 nm - 1300 nm and the peak power densities for this system were of order 10 GW/cm\textsuperscript{2} . At each point in the pump wavelength sweeps the laser energy was adjusted to compensate for the wavelength dependent OPA efficiency, and a separate responsivity value was used. Any residual SH beam from the laser or OPA was filtered out using a combination of dielectric band-pass and color glass filters placed before the sample. Neutral density filters and a pair of crossed polarizers adjusted the pump power, which was TM-polarized. A pair of calcium fluoride lenses focused the TM-polarized pump and collected the TM-polarized SH signal.

The samples were mounted on a custom goniometer with six degrees of freedom and the angle of incidence could be adjusted with a precision of  $0.2^{\circ}$ using a motorized rotation stage. The sample was placed between two different colored glass filters to avoid possible harmonic signals arising from portions of the setup other than the sample. Additional band-pass filters were placed just in front of the detector. By choosing these filters, the total optical density exceeded the efficiency of harmonic generation, making sure that only SH radiation form the Al sample arrives at the detector. Whenever dielectric band-pass filters for the harmonic beam wavelength were available, they were used to confirm that the detected signal was solely related to the harmonic. A mirror on a flip mount allowed the selection of either a calibrated Si photodiode or a sensitive photomultiplier tube (PMT), to enable calibrated measurements of the system responsivity. We expect values of pump to SH conversion efficiencies as low as $ \sim 10^{-13} $, requiring a very sensitive detection system and a detailed calibration procedure to estimate the efficiency of the process as accurately as possible. This calibration uses either transmission or reflection coefficients of each and every optical element placed after the sample, for an accurate calculation of the energy transfer from the pump to the SH beam. The electronic signal from the PMT was filtered and amplified by a low-noise preamplifier before being measured with a digital oscilloscope connected to a computer. The entire detection system is mounted on a rigid platform, which in turn is placed on a rotating arm to allow measurement of either transmittance or reflectance. 

\section{Results and discussion}\label{sec4}

\begin{figure*}[!ht] 
\centering
\includegraphics[width=\textwidth]{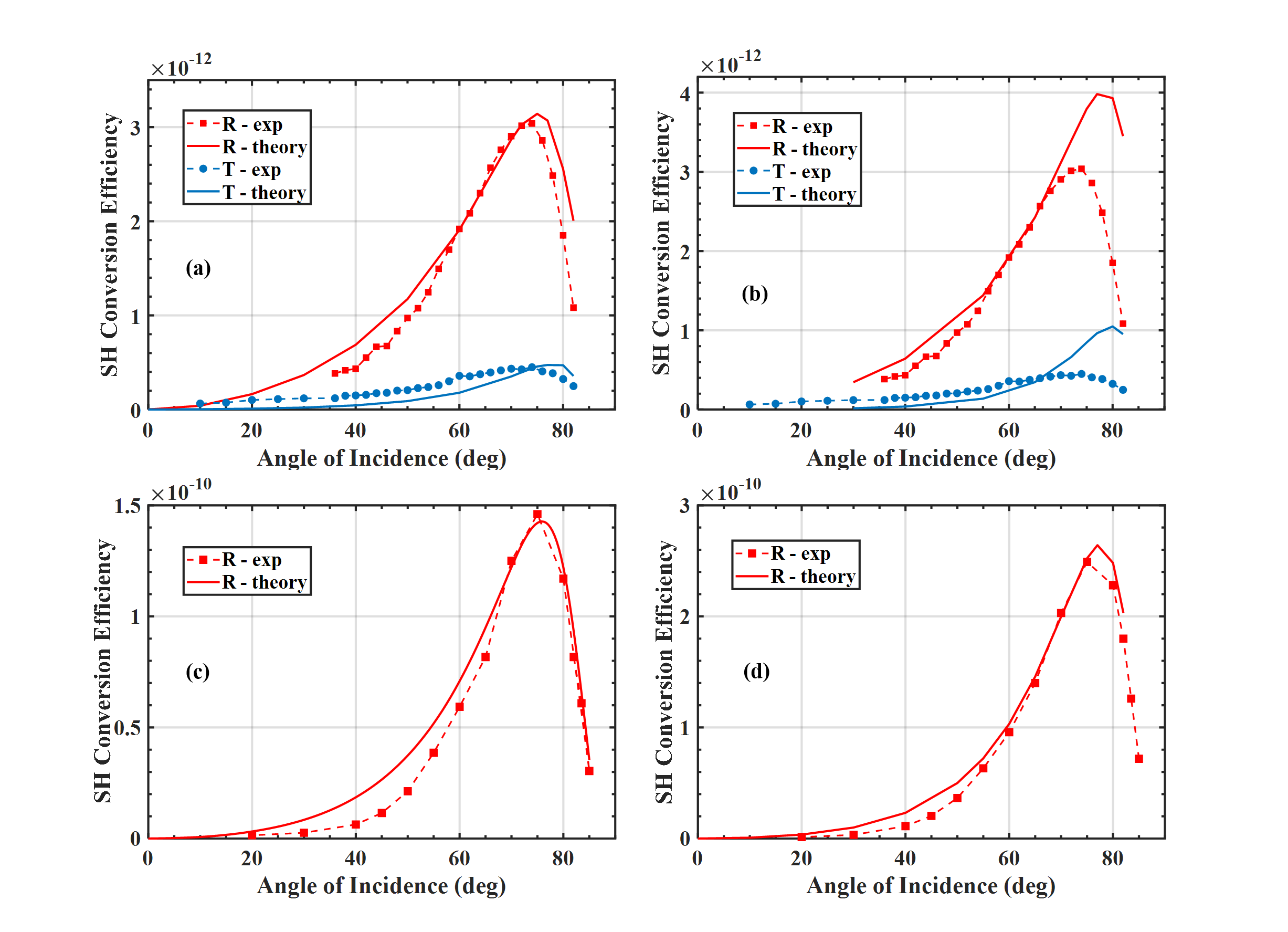}
\caption{(a) Reflected and transmitted SHG conversion efficiency vs. incident angle for 20nm-thick Al layer deposited on a glass substrate. The calculations are carried out assuming Al is covered by a 3nm-thick oxide layer and is grown on a glass substrate having  $\epsilon =2.4$. (b) Same data as in Fig.3(a). The calculations are performed without substrate and oxide layers, showing the importance of properly accounting for all boundaries. Reflected SHG efficiency vs angle for (c) 20nm- (d) and 40nm-thick Al layer, and 900nm carrier wavelength. Incident pulses are assumed to be approximately 100fs in duration with peak power density of 5 GW/cm\textsuperscript{2} in (a) and (b), and 10 GW/cm\textsuperscript{2} in (c) and (d). Measured results are reported with empty markers, while simulations are reported with solid curves.}
\label{fig3}
\end{figure*}

\begin{figure*}[!ht] 
\centering
\includegraphics[width=.9\textwidth]{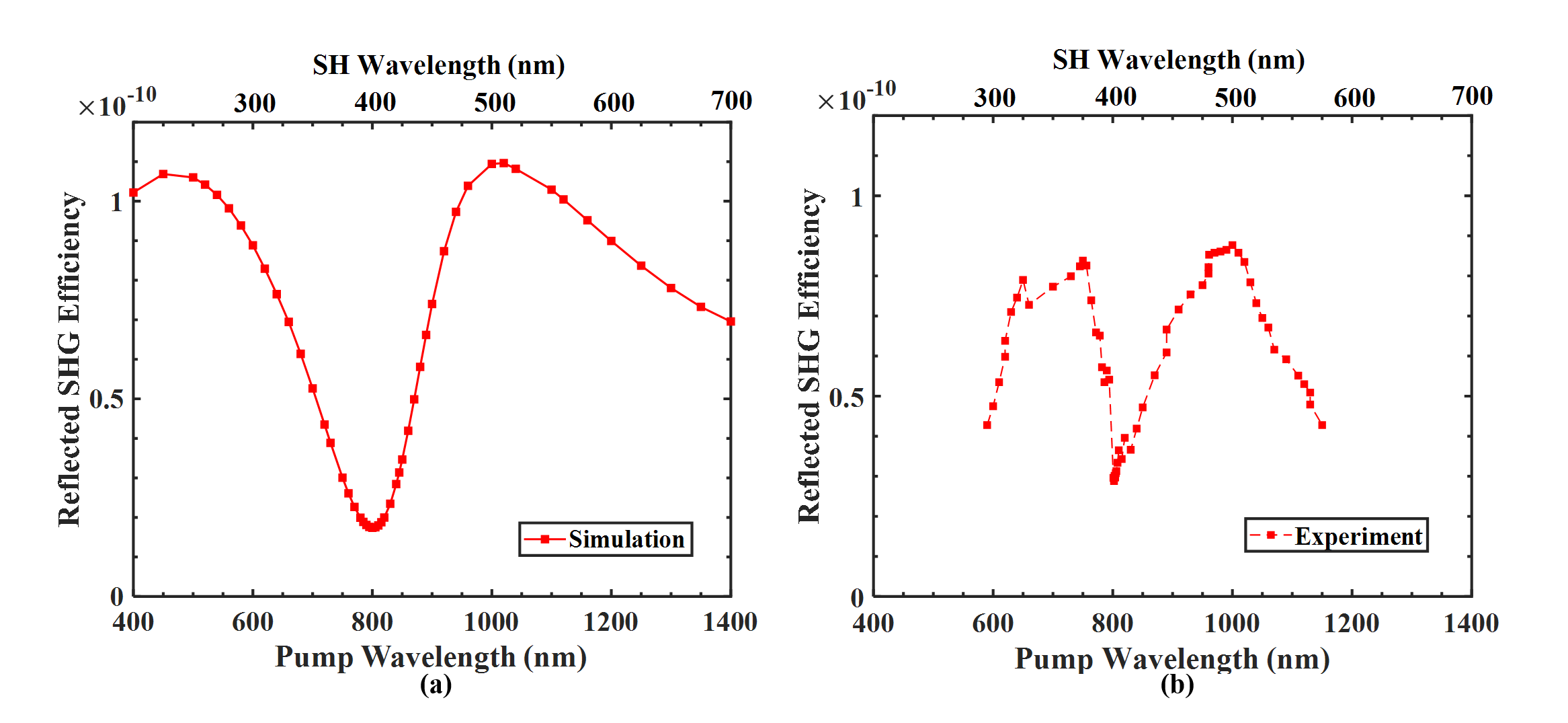}
\caption{(a) Predicted reflected SHG conversion efficiencies from a 40 nm-thick aluminum layer arranged as in the inset of Fig.\ref{fig2}. Pulse duration is 100 fs, incident angle is 75$^\circ$, peak power density approximately 10 GW/cm\textsuperscript{2}. (b) Experimentally retrieved reflected SHG for the same pumping conditions outlined in (a). Both simulation and experiment display a minimum near 800 nm, with remarkable degree of agreement.}\label{fig4}
\end{figure*}

In Fig.\ref{fig3}(a) we report predictions and experimental observations of the angular response of the reflected and transmitted SHG from a 20 nm-thick layer for incident 100 fs pulses tuned to 800 nm, near the peak of absorption. For this layer thickness, the presence of both substrate   substrate and front oxide layer in the simulations improves the agreement with our measurements at large angles because transmittance and feedback are not negligible [compare Figs. \ref{fig3}(a) and (b)]. Similar results are obtained for the reflected SH signal from a 40nm-thick layer (not shown) as both transmitted pump and harmonic signals decrease significantly.  In Figs. \ref{fig3}(c) and \ref{fig3}(d) we depict reflected SHG efficiencies for 20 nm- and 40 nm-thick layers when the carrier wavelength of the incident pulse is tuned to 900 nm, where the aluminum layer is practically a mirror. The simulations account for surface (Coulomb and convection) and magnetic (Lorentz force) sources in the material equations of motion in the free electron response, and similarly in the bound electron equation but with no convective components \cite{bib24,bib25,bib26,bib27,bib28,bib29,bib30,bib31}, given that bound electrons never leave their atomic sites. We limit ourselves to reproducing the equations of motion for free and bound electrons, respectively: 
\begin{widetext}
    \begin{equation} 
\begin{split}
{\boldsymbol{\Ddot{P}}}_f + \gamma_f {\boldsymbol{\Dot{P}}}_f & = \frac{n_{0f} e^2 {\lambda_r}^2}{m_f^* c^2} \boldsymbol{E} - \frac{e \lambda_r}{m_f^* c^2} (\nabla \boldsymbol{\cdot} {\boldsymbol{P}}_f) \boldsymbol{E} + \frac{e \lambda_r}{m_f^* c^2} {\boldsymbol{\Dot{{P}}}_f} \times \boldsymbol{H} \\
& - \frac{1}{n_{0f} e \lambda_r} [(\nabla \boldsymbol{\cdot} {\boldsymbol{\Dot{P}}}_f) {\boldsymbol{\Dot{P}}}_f + ({\boldsymbol{\Dot{P}}}_f \boldsymbol{\cdot} \nabla) {\boldsymbol{\Dot{P}}}_f] + \frac{2}{5} \frac{E_F}{m_0^* c^2} [\nabla (\nabla \boldsymbol{\cdot} {\boldsymbol{P}}_f) + \frac{1}{2} \nabla^2 {\boldsymbol{P}}_f]  
\end{split}
\label{eqn:2}
 \end{equation}
 
 \begin{equation}
{\boldsymbol{\Ddot{P}}}_b + \gamma_{0b} {\boldsymbol{\Ddot{P}}}_b + \omega^2_{0b} \boldsymbol{P}_b + \beta \boldsymbol{P}^3_b
 = \frac{n_{0b} e^2 \lambda_0^2}{m_b^* c^2} \boldsymbol{E} + \frac{e \lambda_r}{m_b^* c^2} (\boldsymbol{P}_b \boldsymbol{\cdot} \nabla)\boldsymbol{E} + \frac{e \lambda_r}{m_b^* c^2} {\boldsymbol{\Dot{P}}_b \times \boldsymbol{H}}
\label{eqn:3}
\end{equation}

\end{widetext}

\begin{figure*}[!ht] 
\centering
\includegraphics[width=\textwidth]{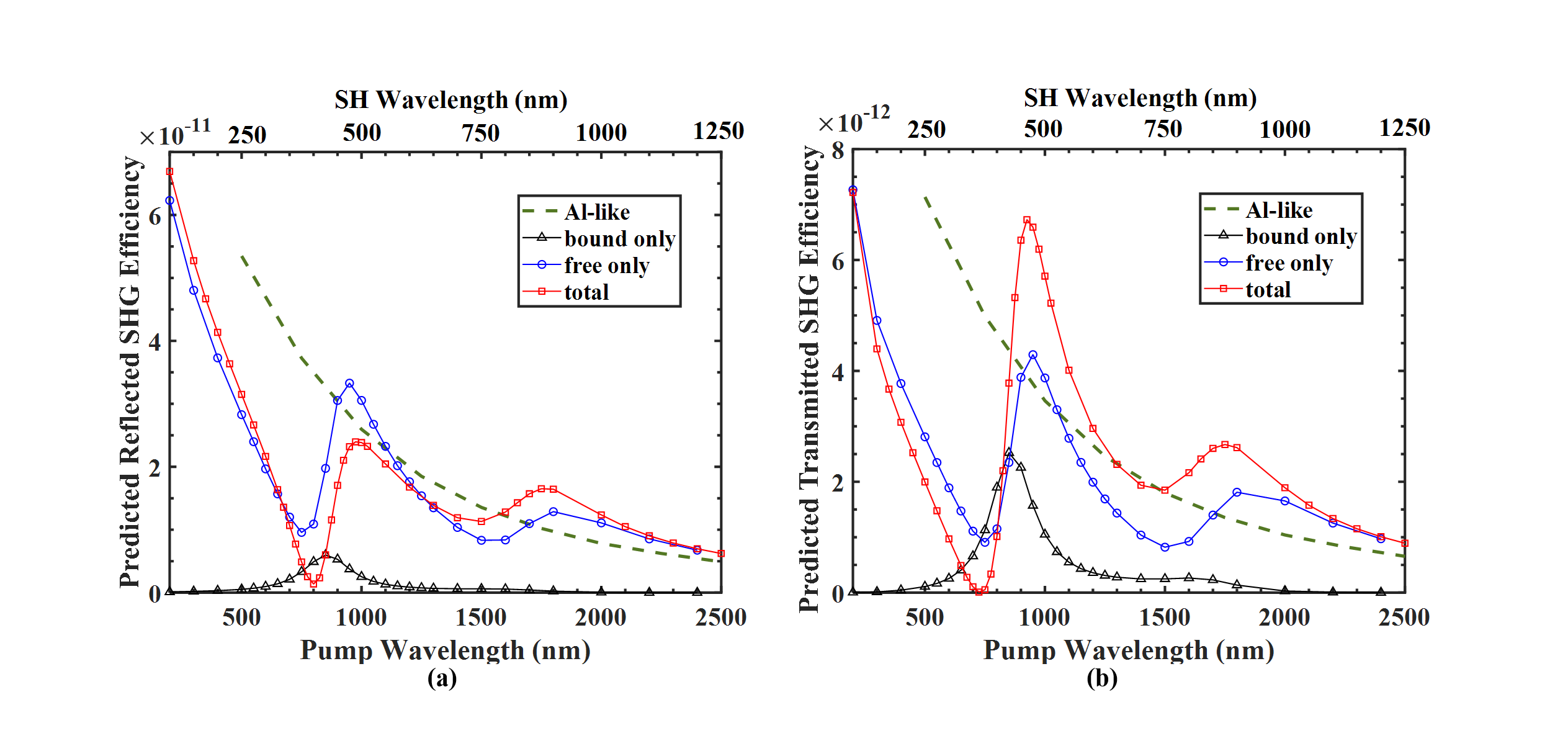}
\caption{(a) Reflected SH conversion efficiencies from a free-standing 20 nm-thick aluminum layer. 100 fs pulses are incident at 60$^\circ$, with peak power density of approximately 10 GW/cm\textsuperscript{2}. Conversion efficiency has been calculated assuming: (i) bound electrons only (black curve), (ii) free electrons only (blue curve), (iii) total contribution coming from both free and bound electrons (red curve), (iv) reflected SH efficiency from an Al-like plasmonic material with a pure Drude response (green, dashed curve). (b) same as (a) for transmitted SH efficiencies.}\label{fig5}
\end{figure*}

Here, $m^*_f$ and $m^*_b$ are the effective free and bound electron masses, and $n_{0f} \approx 18\times 10^{22}/cm^3$ is the free electron density; $\lambda_r = 1\mu m$ is a reference wavelength that scales the spatial coordinates, $ \zeta=\frac{y}{\lambda_r }, \xi = \frac{z}{\lambda_r }$ , where y and z are transverse and longitudinal coordinates, respectively, and $\tau = \frac{ct}{\lambda_r}$  is the scaled time; $E_F$ is the Fermi energy; and $c$ is the speed of light in vacuum. Spatial and temporal derivatives are performed with respect to these new scaled spatial and temporal coordinates. In this wavelength range the amplitude of the dielectric function of Al is of order 100. As a result, nonlocal effects (pressure and viscosity, the two terms in brackets in the last term in Eq.\ref{eqn:2}) perturb only slightly the local dielectric constant. We note that, upon thorough review of the published literature, we have not come across any instance (aside from our own research) that adopts the dynamics outlined by Eqs.\ref{eqn:2}-\ref{eqn:3} under any scenarios. This includes consideration of surface and magnetic contributions within the bound electron components, the last two terms on the right-hand side of Eq.\ref{eqn:3}, which appear to be crucial not only in the vicinity of the resonance but also in spectral regions where one might expect their influence to have faded.

The $a$, $b$, and $d$ coefficients reported in Eq.\ref{eqn:1} are usually derived solely from manipulations of Eq.\ref{eqn:2} (free electrons only), the constitutive relations, and the assumption of relatively thick layers \cite{bib4,bib6,bib24}. The presence of surface and magnetic terms in the bound electron Eq.\ref{eqn:3} adds crucial nonlinear second order sources absent in the theory that interfere with free electron sources. Once the dielectric function is fitted as in Fig.\ref{fig1}, the only remaining free parameters in Eqs.\ref{eqn:2}-\ref{eqn:3} are the free and bound electron masses, to some extent the free electron density, which are adjusted depending on the incident peak power density, and slight thickness variations. For example, for the 20 nm layer and the pump tuned to 900 nm we used the combination $m^*_f \approx \frac{m_e}{2}$; $m^*_b \approx \frac{m_e}{2}$; $n^*_{f} \approx \frac{n_{0f}}{2}$, and $m^*_f \approx \frac{m_e}{3}$; $m^*_b \approx \frac{m_e}{3}$; $n^*_{f} \approx \frac{n_{0f}}{3}$ for the 40nm-thick film. This flexibility is justified because it is well-known that the density and dielectric constant of sputtered samples can change with thickness, with decreased density and increased porosity near the surface exposed to the vacuum, while the opposite occurs on the substrate side \cite{bib32}, with possible consequences to the parabolicity of the band. As a result, the assumption that the effective masses are constant throughout the sample and across the entire wavelength range is likely an oversimplification. While these choices do not change the respective plasma frequencies for each sample (the ratios $\frac{n}{m}$ remain constant), and therefore induce identical dielectric responses in both layers, SHG conversion efficiencies are directly proportional to incident peak power density, and to $\frac{1}{m^*_{f,b}}$  and $\frac{1}{n_f}$ , as may be ascertained from Eqs.\ref{eqn:2}-\ref{eqn:3}. Allowing for slight effective mass and density changes across the spectral range leads to remarkable agreement between simulations and measurements. The results in Fig.\ref{fig3} suggest that SHG efficiencies change appreciably at large angles if oxidation layer and substrate are excluded from the simulations. However, the qualitative aspects are preserved, because spatial derivatives of the fields and polarizations, and surface and bulk currents do not change appreciably, given the large dielectric constant of bulk aluminum relative to either oxide, vacuum, or substrate.

In Fig.\ref{fig4}(a) we display our predicted spectral response of reflected SHG efficiency across visible and IR wavelengths. In Fig.\ref{fig4}(b) we show the measured SHG across the same range. We have separated the figures to capture both qualitative and quantitative aspects of the results, by maintaining constant effective masses and density in the calculations. While the locations of the theoretical and measured minima coincide in both figures, the measured spectral response is somewhat sharper compared to the simulations. This may be due to our imperfect knowledge about the sample (primarily varying charge density and effective masses as a function of depth), and the assumption that the effective mass is constant across the entire range investigated. This notwithstanding, the simulated results and experimental observations look remarkably similar and display the same minimum near resonance. As before, the only free parameters used in the simulations of Fig.\ref{fig3}-\ref{fig4} are free and bound electron masses. Additional calculations also suggest that the location of the minimum does not change appreciably for different layer thicknesses.

In Fig.\ref{fig5} we consider and examine the interplay between free and bound electrons in the context of SHG, an interaction that will also be pivotal for higher harmonics for materials other than aluminum. For instance, Au, Ag, and most semiconductors and conductive oxides display a Lorentzian response in the UV range, where it then becomes essential to incorporate material dispersion accurately. Each figure contains three curves that represent either total reflected [Fig.\ref{fig5}(a)] or transmitted [Fig.\ref{fig5}(b)] SHG efficiency, and then either SHG by free or bound electrons only, obtained for 20 nm-thick layers by sequentially zeroing out either the nonlinear terms in either the bound or free electron equation, respectively. Similar curves are obtained for thinner or thicker nanolayers regardless of substrate choice. The results show that when only the nonlinearities in the bound electron equation are considered, SHG efficiencies display a peak at the absorption resonance, while SHG from free electrons-only displays a relative minimum near 810 nm. In contrast, total SHG efficiency shows an absolute minimum when both free and bound electron nonlinearities are included. This is precisely where a comparison was made between silver and aluminum by assuming a mostly free electron system \cite{bib4}, and where it was concluded that Ag was a more efficient frequency converter. It is remarkable that while free electrons impart the generic spectral shape to the response, both reflected and transmitted efficiencies from bound electrons are of the same order of magnitude as the free electron-only efficiencies, and in fact contribute to the extent that they interfere destructively and nearly completely suppress the SH signals precisely near resonance. The total SH reflected efficiency may be estimated as follows: 

\begin{equation}
    R^{total}_{2w} \approx R^{free-only}_{2w} - 2R^{bound-only}_{2w}
    \label{eqn:4}
\end{equation}

This confirms the point that bound electron contributions tend to add destructively to the process. In Fig.\ref{fig5} we also plot the outcome of a calculation where the Lorentzian component of the dielectric response has been arbitrarily set to zero, so we are dealing with a pure Drude response. Both reflected and transmitted SHG efficiencies increase monotonically with decreasing wavelength. This is in contrast with SHG from gold layers modeled with their actual material dispersion, where SHG efficiency decreases with decreasing wavelength \cite{bib28}. Finally, we note that the predicted peak near 1600 nm in the SH spectrum is due to the SH signal experiencing anomalous dispersion across the absorption resonance, an effect clearly missing in the Drude-only response. Therefore, taken together, these facts amount to clear evidence that in Al bound electrons completely determine the dynamics up and down the electromagnetic spectrum, and strongly suggest that surface phenomena should be reevaluated accordingly depending on the material under consideration.

\section{Conclusions}\label{sec5}

In view of our predictions and observations, we conclude that the influence of bound electrons on SHG spans a very broad frequency range from the UV to the NIR, either by dominating near resonance, or by acting as a catalyst that facilitates energy exchange below 500 nm and all the way up to approximately 2000 nm. The most likely explanation for the behavior shown in Fig.\ref{fig4} and Fig.\ref{fig5} is that near resonance most of the incoming pump energy is absorbed by the bound electrons, which respond linearly and nonlinearly, effectively screening and reducing the energy available to free electrons. The surprising aspect here is that away from the resonance peak, for example near 1600 nm, where bound electrons do not by themselves yield much SH signal, they nevertheless act like a catalyst as the energy that flows through them opens an exchange channel that can significantly impact total conversion efficiencies. The repercussions on both linear and nonlinear light-matter interactions are thus significant beyond SHG and beyond aluminum, as the influence of bound electrons makes itself felt across the visible and IR spectrum, including at wavelengths where one might expect only a pure plasmonic response. More generally, it is evident that the inadequacy of the free electron model for one material requires reassessment and novel approaches to fully account for the most salient aspects of linear and nonlinear surface interactions of other materials, including doped semiconductors, conductive oxides, and time varying media. In the latter case, it is especially important to first and foremost recognize that in the kind of pump-probe systems that are currently under investigation \cite{bib33}, exceptionally strong ultrafast pumps can completely alter the spatio-temporal dynamics of the free electron density in a way that will require the inclusion of additional terms in Eqs.\ref{eqn:2}-\ref{eqn:3}. For instance, a spatial variation of the free electron density will trigger additional linear, nonlinear, nonlocal, and convective sources, as well as higher order contributions to account for higher harmonics and frequency down-conversion \cite{bib25,bib26,bib27,bib28,bib29,bib30,bib31,bib34,bib35}, as boundaries are spatially and temporally modulated first by the pump, and then by the probe. Therefore, at a minimum, care should be exercised when simple ad-hoc explanations are adopted, and when conclusions are drawn without the benefit of a suitable theoretical model.

\section{ACKNOWLEDGEMENTS}\label{sec6}

SM, JT, and CC acknowledge Spanish Agencia Estatal de Investigación (project no. PID2019-105089GBI00/AEI/10.130397501100011033) and US Army Research Laboratory Cooperative Agreement N$^\circ$ W911NF-22-2-0236 issued by US ARMY ACC-APG-RTP. M. A. V. and D. d. C. thank NATO SPS Grant no. G5984 - RESPONDER for financial support.

\section{References}\label{sec8}
\bibliography{aipsamp}

\end{document}